\documentclass[10pt,twocolumn,twoside]{IEEEtran} % Format final IEEE
\usepackage{amsfonts, latexsym, amsmath}

\usepackage[utf8]{inputenc}
\usepackage[english]{babel}
\usepackage{fixltx2e}
\usepackage{graphicx}
\usepackage{algorithmic, algorithm}
\usepackage[hyphens]{url}

\usepackage{breakurl}
\usepackage{paralist}
\usepackage{tikz}
\usetikzlibrary{decorations.markings}
\usetikzlibrary{trees}
\newtheorem{defin}{Definition}
\newtheorem{prop}{Proposition}
\newtheorem{remark}{Remark}

\newcommand{\opreturn}{\texttt{OP\_RETURN}}

\newcommand{\addr}{\texttt{a}}

\newcommand{\Prover}{\mathcal{P}}
\newcommand{\Verifier}{\mathcal{V}}

\newcommand{\SP}{\mathcal{SP}}
\newcommand{\USR}{\mathcal{USR}}
\newcommand{\IV}{\mathcal{IV}}
\newcommand{\SE}{\mathcal{SE}}
\newcommand{\txenroll}{\texttt{TX}_\texttt{ENROLL}}

\title{Transforming face-to-face identity proofing into anonymous digital identity using the Bitcoin blockchain}

\author{
	\IEEEauthorblockN{Daniel Augot \IEEEauthorrefmark{1} \IEEEauthorrefmark{2} \IEEEauthorrefmark{3}, Herv\'e Chabanne \IEEEauthorrefmark{4} \IEEEauthorrefmark{5}, Olivier Cl\'emot \IEEEauthorrefmark{4}, William George \IEEEauthorrefmark{2} \IEEEauthorrefmark{1} \IEEEauthorrefmark{3}}
	\\ \IEEEauthorblockA{\IEEEauthorrefmark{1}INRIA, Palaiseau, France 
	}\\
	\IEEEauthorblockA{\IEEEauthorrefmark{2}Laboratoire LIX, \'Ecole Polytechnique \& CNRS UMR 7161, Palaiseau, France
	}\\
	\IEEEauthorblockA{\IEEEauthorrefmark{3} Universit\'e Paris-Saclay, Palaiseau, France
}\\
	\IEEEauthorblockA{\IEEEauthorrefmark{4}OT-Morpho, Issy-les-Moulineaux, France
	}\\
	\IEEEauthorblockA{\IEEEauthorrefmark{5}T\'el\'ecom ParisTech, Paris, France
	}
}

\IEEEoverridecommandlockouts
\begin{document}

\date{}
\maketitle

\begin{abstract}

The most fundamental purpose of blockchain technology is to enable persistent, consistent, distributed storage of information. Increasingly common are authentication systems that leverage this property to allow users to carry their personal data on a device while a hash of this data is signed  by a trusted authority and then put on a blockchain to be compared against. For instance, in 2015, MIT introduced a schema for the publication of their academic certificates based on this principle. In this work, we propose a way for users to obtain assured identities based on face-to-face proofing that can then be validated against a record on a blockchain.  Moreover, in order to provide anonymity, instead of storing a hash, we make use of a scheme of Brands to store a commitment against which one can perform zero-knowledge proofs of identity. We also enforce the confidentiality of the underlying data by letting users control a secret of their own. We show how our schema can be implemented on Bitcoin's blockchain and how to save bandwidth by grouping commitments using Merkle trees to minimize the number of Bitcoin transactions that need to be sent. Finally, we describe a system in which users can gain access to services thanks to the identity records of our proposal.

\end{abstract}


\section{Introduction}

It is a common occurrence in modern life to authenticate part of one's identity by appealing to an existing relationship with a third-party service. Users have primary forms of identification such as passports and drivers' licenses issued by governments, but they may also need to use secondary forms of identification from trusted sources whose business model nevertheless is not that of an identity provider. For example, in order to establish one's address before buying a cellular plan or obtaining a library card, one can provide a utility bill. One might want to replace this system by one in which a digital record is issued to users that can serve the role of the paper utility bill. %facilitating this exchange  even as we allow users to only show the minimum amount of information required to authenticate. 
A simple way of doing this is for the identity issuer to sign a digital identity document that the user can then store on her device. However, in such a system, it can be difficult for the issuer to update or revoke the document if there are changes to the user's identity. Issuers can create revocation lists, such as the ones used in public key infrastructures, but using these lists can pose subtle challenges \cite{revocationlists}. We will propose a system for this setting that uses properties of blockchains to provide for streamlined, integrated revocation and updating of identities. 

\section{Related work}

There are several proposals to use blockchains to store identity information. For a survey of such proposals see \cite{BlockIDSurvey} and \cite{BlockStartupSurvey}. In particular, MIT Media Labs \cite{MIT} proposes a system in which academic certificates can be verified against records stored in the Bitcoin blockchain and which are considered revoked if the issuer spends a Bitcoin transaction output. The Blockstack project \cite{Blockstack} has implemented decentralized versions of PKI and DNS using the Bitcoin blockchain. The uPort \cite{uPort} project, which has been implemented in the Ethereum \cite{Ethereum} blockchain, includes a smart contract that allows a user who has lost her device to regain access to an identity after being authenticated by several pre-designated contacts.  
Other proposals involve new blockchains that are designed for their specific applications. Namecoin \cite{Namecoin} and Certcoin \cite{Certcoin} have implemented DNS and PKI respectively on the Namecoin blockchain. In \cite{JH}, a system is proposed to use zero-knowledge proofs and either Namecoin or Bitcoin to issue identity credentials in settings such as anonymous peer-to-peer networks, where one does not have trusted credential issuers. While using an application designed blockchain can provide greater flexibility, they have less mining power than Bitcoin, which can have security implications for schemes built on top of them, see the security analysis of the Namecoin blockchain in \cite{Blockstack}. Similarly, IDCoins \cite{IDCoins} uses a custom blockchain whose proof of work is related to the generation of GPG/PGP keys. These keys are then used to create a web of trust. In \cite{Zyskind} a system is proposed to store user personal information such as the GPS data from their phone in a distributed hash table; then this hash table is coupled with a blockchain that stores pointers to the data and permissions on how it may be used. The Estonian government has built an electronic records system based on the Guardtime KSI blockchain \cite{eestonia}, \cite{estoniannews}, which is a permissioned, namely only authorized parties can publish transactions to this blockchain. The proposition of ChainAnchor \cite{ChainAnchor} attempts to create a semi-permissioned structure on existing blockchains, potentially including the Bitcoin blockchain, by changing the incentive structure of miners to promote transactions that have passed a layer of authentication. 

\section{Our contribution} 

We present a identity management system based on the Bitcoin blockchain that allows for a very flexible user experience, providing identity documents that can be used in a wide variety of use cases. To achieve this flexibility, it is necessary to provide protections for user privacy, particularly considering the public nature of information on blockchains. To do this, we incorporate into our protocol a zero-knowledge, selective disclosure identity scheme due to Brands \cite{Brands}. This scheme corresponds particularly well to our proposal as its security is based on cryptographic primitives that are already used in Bitcoin. % and as the commitments against which one selectively discloses are small enough to be easily stored in Bitcoin transactions.
Moreover, by combining the potential to only reveal the information about one's identity required for a given authentication with identity records stored on the global, widely accessible Bitcoin network, we empower users to take greater control over their identities. 

Compared to most previous work that uses blockchains to offer comparably general identity documents \cite{eestonia}, \cite{IDCoins}, we consider it an advantage that we work within the structure of the existing Bitcoin blockchain, the most established and secure blockchain and the blockchain that is supported by the greatest amount of mining power, in a way that is consistent with Bitcoins' existing incentive structure. By doing so, our system has the same security as Bitcoin itself while minimizing the additional infrastructure required. Compared to \cite{JH}, which explores the possibility of using either Bitcoin or Namecoin, we further develop ways in which one can embed aspects of identity management into the transactional structure of Bitcoin. 

%As we will see that the cryptographic primitives of our systems (hash functions, discrete logarithms) are already incorporated into Bitcoin, our system  %too difficult to implement. 

%Moreover, our proposal allows identity issuers, such as the utility company, to maintain less user information in storage even as customer identity can still be confirmed. As laws regulating the storage of personal data become more stringent, such as under the soon to come into effect General data protection regulation of the European Union \cite{euprivacy}, storing less customer data can mean less investment required in compliance departments and lower financial liabilities for data breaches. 

Specifically, we build on the ideas of using this transactional structure to encode revocation of certificates in \cite{MIT} and updating of logs in \cite{Catena}. These methods allow us to provide for an integrated, streamlined mechanism for revocation and updating of the identity documents issued by our system. In doing so, we resolve several technical challenges related to the interplay between revocation, the structure of Bitcoin, and Brands' anonymous credentials. Particularly, we provide a means for the user to have a secret required for authenticating against an identity record that continues to perform this role even as the user's identity record is updated. %that can be updated automatically when the user's identity record is updated in a way that preserves confidentiality.

%identity issuer can update

%We will detail in Section \ref{sec:Enroll} that, in our scheme, identity verifiers will commit a user's identity to the Bitcoin blockchain. This will have the advantage of providing a neutral space that is persistent in time from which the user can demonstrate her identity to diverse service providers. Moreover, Bitcoin protection against double spending can be used as a mechanism to allow identity verifiers to update, revoke, or attach a notification to a user's identity in non-equivocable ways. This structure is similar to that of Catena \cite{Catena} and to how revocation is performed via a semantically encoded Bitcoin transaction in the academic
%certificates scheme of MIT Media Lab \cite{MIT}.  

%Our proposal, by 

\section{Background}

%\begin{defin} An identity is a tuple
%	$(X_1, \ldots, X_n)$ where each $X_j\in \mathbb{Z}_q$ stands for a different
%	attribute, as exemplified below.
%\end{defin}

Our system will involve commitments to users' identities being written into the Bitcoin blockchain. The commitment scheme that we make use of, due to Brands \cite{Brands}, will allow users to disclose selective elements of their identity against this commitment in keeping with the principle that only necessary information about a user should be circulated and stored. Also, we will notice that Brands selective disclosure uses similar cryptographic primitives to Bitcoin, discrete logarithms and hash functions, so the security of our two building blocks are related. %As the Bitcoin blockchain is public, this necessitates  

\subsection{Brands selective disclosure scheme} \label{back:Brands} 

%We will make use of a system for selectively disclosing elements of an identity due to Brands, \cite{Brands}. We briefly recall this system, which relies on discrete logarithms and hash functions.  

%All the following is
%from~\cite{Brands}. 

%Suppose we want to make selective disclosures involving an identity $(X_1, \ldots, X_n)$. 
We follow \cite{Brands}. Suppose we want to make selective disclosures involving an identity with $n$ fields, $(X_1, \ldots, X_n)$. (For example, $X_1$ may represent a user's name, $X_2$ her nationality, etc). Let $q$ be a prime
number and $G$ a group of order $q$, in which the discrete logarithm is
hard. For our purposes, we will take $G$  to be the Koblitz elliptic curve secp256k1, (we use multiplicative notation for compatibility with \cite{Brands}). Note that by using the same group that is used in the Bitcoin signature protocol, we reduce the number of different cryptographic primitives on which our system depends. %Note that this is the same group used for %We choose this group as it is also used for 
%the Bitcoin signature protocol. 
Let $g_0,g_1,\ldots,g_n \in G$.

%In~\cite{Brands}, Brands proposed very efficient ways of
%revealing parts of an identity to verifiers, relying on discrete
%logarithms (he also proposed a system based on RSA, however here we restrict ourselves to DL) and hash functions. All the following is
%from~\cite{Brands}. 
%Assume that $n$ identity fields $X_1,\dots,X_n$ are to be
%cryptographically blinded  for further proofs. Let $q$ be a prime
%number and $G$ a group of order $q$, in which the discrete logarithm is
%hard. Typically, we take $G$  to be the Koblitz elliptic curve secp256k1 where points are represented with 64 bytes (we use multiplicative notation for compatibility with \cite{Brands}), namely we use the same $G$ that is already being used for the Bitcoin signature protocol. Let $g_0,g_1,\ldots,g_n \in G$.

Furthermore, Brands notes that there is the need for an
auxiliary random $X_0$. This will prevent an attacker who knows some of the $X_j$ fields for a user from performing a dictionary attack in which she guesses values for the other $X_j$. %See Sections \ref{sec:Enroll} and \ref{sec:update} for further information on the role that $X_0$ will play in ou% protect unknown fields from a dictionary attack when the other fields are known. 

\begin{defin}
	The tuple $(X_0,X_1,\ldots,X_n) \in \mathbb{Z}_q^{n+1}$ is called a
	Discrete Logarithm REPresentation (DLREP) of
	$h=\prod\limits_{j=0}^{n} g_j^{X_j} \in G$ with respect to
	$(g_0,g_1,\ldots,g_n)$.
\end{defin}

In order for a prover $\Prover$ to establish knowledge of a DLREP of $h$ to a verifier $\Verifier$, the following procedure is performed %a
%prover $\Prover$ performs the following procedure
~\cite[\textsection 2.4.3]{Brands}
\begin{enumerate}
	\item $\Prover$ generates $n+1$ random, secret numbers $a_0,a_1,\ldots,a_n$. Let $A=\prod\limits_{j=0}^{n} g_j^{a_j}$. $\Prover$ sends $A$ to $\Verifier$. \item $\Verifier$ provides a challenge number $c$.   
	%\item $\Prover$ generates $n+1$ secret, random numbers $a_0,a_1,\ldots,a_n$ in
	%$G$. Let $A=\prod\limits_{j=0}^{n} g_j^{a_j}$, and compute $c$ as $c=\Hash(A)$, 
	%where $H$ is a one-way
	%hash function. 
	\item $\Prover$ computes $b_j=a_j+c X_j$, $j=0,1,\ldots,n$ and sends them to $\Verifier$.
	\item The verifier $\Verifier$  checks that $\prod\limits_{j=0}^{n} g_j^{b_j} h^{-c}=A$ holds.
	%$\Hash(\prod\limits_{j=0}^{n} g_j^{b_j} h^{-c})=c$ holds.
\end{enumerate}
We denote such an interactive proof $\pi$.
%\begin{prop}\cite[Proposition 2.4.8]{Brands}
%  The above protocol is complete and perfectly
%  witness-indistinguishable, regardless of the process of generating
%  $\Verifier$'s challenge (and thus 1' above is correct).
%\end{prop}
%Then, \cite[Chapter 3]{Brands} shows how the Discrete
%Logarithm REPresentation can be used to selectively disclose
%proof of properties about the $X_j$'s, while any other information remains
%hidden. 
Note that it is essential that $\Prover$ knows all of the $X_j$ in
order to be able to perform step 3. Particularly, $X_0$ which is
secret and generated randomly acts as a sort of key to be able to
perform these proofs.

More generally, in \cite[Chapter 3]{Brands}, it is shown how to prove arbitrary satisfiable Boolean statements about the
$X_j$'s using such a Discrete
Logarithm REPresentation without revealing other information. Thus, there is a great deal of flexibility in what a user can prove about her identity. She can, for example, provide a proof that she is a French citizen AND that she lives at a certain address, or that she is under 18 OR over 65. In this way, the user can prove (true) statements about her identity that contain an arbitrary number of ANDs, ORs, and NOTs in such a way that a verifier learns only the content of the statement. Specifically, Brands shows:  %See Appendix \ref{append:size} for further details and estimates on the 
%size of the proofs required. Brands shows:

\begin{prop}\cite[Proposition 3.6.1]{Brands} \label{prop:Brands}
	There is a constructive protocol for demonstrating arbitrary satisfiable Boolean formulas in which $X_0$ does not appear such that: \begin{enumerate} \item The protocol is complete and sound \item The protocol is a proof of knowledge of a DL-representation $h$ with respect to the tuple $(g_0,\ldots, g_n)$ \item For any distribution of $(X_1,\ldots, X_n)$, whatever information a verifier in an adaptively chosen formula attack can compute about $(X_1,\ldots, X_n)$ can also be computed using merely its a priori information and the status of the formulas requested.  \end{enumerate} 
	
\end{prop}

Moreover, Brands~\cite{Brands} shows that if the discrete logarithm problem is
difficult, DLREP is one-way and collision-intractable, preventing forgery attacks.

We make two remarks on subtleties involved in the use of this protocol: 

\begin{remark}
	Note that $\Prover$ should not re-use the same $A$ with different challenges $c$, see Section 5.4 of \cite{Brands} on limited use identities; however, one may prove multiple different formulas about the same DLREP $h$ if the $A$ is changed for each challenge.
	
\end{remark}

\begin{remark} \label{rem:dict}
	One should be careful to not publish two commitments using the same $X_0$. Namely, one should not publish $h=\prod\limits_{j=0}^{n} g_j^{X_j}$ and $h'=\prod\limits_{j=0}^{n} g_j^{X'_j}$, with $X_0=X_0'$,  as this would allow potential attackers who can guess plausible values for (some of) the $X_j$ and $X'_j$ to calculate $h/h'=\prod\limits_{j=1}^{n} g_j^{X_j-X'_j}$, which no longer contains a blinding factor and might permit a dictionary attack. 
	\end{remark} 

We will see in Section \ref{sec:Enroll} that the issue discussed in Remark \ref{rem:dict} presents difficulties in issuing an identity which can be updated by an identity issuer even as the user maintains $X_0$ secret. We will overcome this problem by splitting $X_0$ into two pieces of information, both of which are necessary to perform the selective disclosure proofs: $X_{00}$, which is only known by the user, and $X_{01}$, which can be updated by the issuer, see Sections \ref{sec:calc}, \ref{sec:update}.

%be no other information leakage than what the value formula leaks in
%essence about the values $X_1,\dots,X_n$.

\subsection{Bitcoin relevant notions} \label{subsec:Bitcoinsyntax} 
Bitcoin transactions are made up of inputs and outputs; in fact, all bitcoins exist 
%It is a particularity of Bitcoin that all
%bitcoins exist 
in the form of Unspent Transaction Outputs (UTXOs)
\cite{satoshi}, \cite[Chapter 5]{MasteringBitcoin}. A given transaction can have
several inputs, each of which was an output of some previous 
Bitcoin transaction. Each transaction also has one or more outputs each of which has an amount of bitcoin attached to it. For each input, a script must be provided establishing the right to spend the input. (A script which details what must be established is included in the transaction that issued this input as an UTXO.) 
Most transaction outputs correspond to a Bitcoin ``address,'' which is generally the hash of the public key
that can spend that output. Then to spend that output one must merely provide this public key and sign the transaction with the private key. % it or a hash of a script detailing how the coin can 
%be reclaimed. (These are called Pay to Public Key Hash P2PKH and Pay to Script 
%Hash P2SH outputs respectively.) 
A special output type that is relevant to our work is that of $\opreturn$ 
outputs; each such output contains up to $80$ bytes of space in which the sender
of a transaction can store arbitrary information. A requirement of $\opreturn$ outputs
is to have zero bitcoins associated to them; as such, they are provably not spendable, avoiding the necessity for miners to store them as UTXOs. Non-$\opreturn$ outputs must have positive bitcoin amounts assigned to them, and in fact there is a minimum amount per output in order for a transaction to be considered ``standard'' and be included in the blocks of miners who use the Bitcoin Core software \cite{MasteringBitcoin}. This amount varies slightly based on the output, but as of version 0.14 (March 2017) of Bitcoin Core \cite{Dustcode}, it is on the order of $.000005$ bitcoin, which is around $.01$ USD (1 BTC=2280 USD, May 2017).

%The inputs and outputs of a single transaction may
%use the same Bitcoin address repeated several times, or they may use different
%Bitcoin addresses. 
When creating a Bitcoin transaction, a user broadcasts a ``raw transaction'' to the nodes 
containing: the amount of bitcoin to associate to each output, the script 
setting up the requirements to spend each output, and the scripts
for each input that satisfy the requirements established when the corresponding output was created. These requirements typically include signing the transaction with a private key corresponding to the previous output. Each transaction then has a transaction identifier (txid) which is the hash of this raw transaction. A Merkle tree is formed from these transaction identifiers the root hash of which is included in the block header. Hence, the raw transaction, including which inputs and outputs were involved in a given transaction, is committed to in
the blockchain in an immutable way. 

%Particularly, the given inputs and outputs 
%of a given transaction are provably linked together. 

%Miners are compensated by ``fees.'' The amount paid in fees for a given transaction 
%is the difference between the combined values of the
%inputs and the combined values of the outputs. Miners, who are limited in how many bytes they can fit a given block, generally choose to include the transactions with the most profitable fees 
%%%%Note the fee that needs to be paid for a transaction to be included in a block depends 
%with respect to the number of bytes in its raw transaction \cite[Chapter 5]{MasteringBitcoin}. %As such, the fees for each of the transactions that make up our proposal will be distinct.
%These fees will provide an additional overhead cost to our system which we see to be minimal. When discussing our schema,

\section{Actors and protocol structure} \label{sec:actors}

In our system, a user ($\USR$) wishes to authenticate her identity to a service provider ($\SP$). We will have the following additional actors:

\begin{itemize}
	%\item Users ($\USR$)
	\item Identity Verifiers ($\IV$) - organization such as a bank or utility company that has an existing relationship with $\USR$ and can provide justification for aspects of her identity.
	\item Service Enablers ($\SE$) - capable of verifying the records that various $\IV$ have created for $\USR$ in the blockchain and conveying information about these records to $\SP$. The public key $pk_{\SE}$ of $\SE$ should be well-known. %and provides the service of communicating with $\USR$ and $\SP$, can 
	%\item Service Providers ($\SP$)
\end{itemize}

Just like a bank or utility company in a traditional setting, we assume that $\IV$ has a complete knowledge of $\USR$'s identity record. Thus $\USR$ must confer to $\IV$ a great deal of trust to not misuse this personal information. $\IV$ publishes commitments to identity documents into the Bitcoin blockchain on the basis of which $\USR$ can justify her identity, and $\IV$ publishes updates to these records if necessary. 

When required to demonstrate that her identity satisfies some requirements determined by $\SP$ in order to obtain a service, $\USR$ will prove this information to $\SE$. Thus, $\SP$ must trust $\SE$ to accurately relay whether the user meets the requirements and not to perform fraudulent authentications. Moreover, $\SE$ will also learn personal information about the user, namely whatever $\USR$ had to demonstrate to $\SP$ and the fact that $\USR$ obtained a service from $\SP$. (In contrast, $\SP$ will not necessarily learn anything about $\USR$ beyond the fact that a client of $\IV$ that satisfies its requirements obtained a service through the intermediary of $\SE$.) $\USR$ needs to trust $\SE$ to not misuse this personal information; however, we will see that $\USR$ only needs to reveal partial information about her identity to $\SE$. Also, we imagine that there are several companies willing to fill the role of the service enabler, so if $\USR$ is worried that $\SE$ is developing too complete a profile on her, she can perform future authentications through some other service enabler. 

For a further discussion of the assumptions we make on $\IV$ and $\SE$, see the discussion on our security model in Section \ref{sec:security}.

\begin{remark}
	The service enabler has infrastructure allowing it to manage the authentication operations that may be required of a user by a service provider. Some sophisticated service providers may prefer to perform these operations themselves rather than make use of a third party. Similarly, identity verifiers can also perform the role of the service enabler themselves while still retaining the some of the advantages of using a blockchain, particularly in terms of ease of revocation and updates. However, by outsourcing the service enabler role to a specialized entity, $\IV$ can reduce infrastructure costs. Also, $\IV$ does not need to be as ``lively'' as $\SE$. Namely, whereas $\SE$ must be continuously online to enable authentications, $\IV$ only needs to come online periodically to publish or update identities. Also note that if one combines the $\SE$ role with either the $\IV$ or the $\SP$ role, there is a pooling of information from the different roles that the user may not be comfortable with, and the users ability to change $\SE$ periodically for greater anonymity as discussed above would be limited. Thus, users may prefer the model with a distinct $\SE$. 
\end{remark} 

\begin{remark}
	Note that a given user may have accounts with several different $\IV$ (her bank, her utility company, her university, etc) and want to authenticate herself to several different $\SP$. However, she can authenticate through the same $\SE$ for each of these accounts; thus $\SE$ can serve as a single service sign-on. %Additionally, there may be several different companies that are willing to fill the role of $\SE$; then $\USR$ is free to use an identity issued by $\IV$ via any of these service enablers. 
\end{remark}

In practice, we imagine $\USR$ interacting with $\SE$ via a website or
a mobile phone app. The user's identity information will be entered
into her computer or phone, but not disclosed to $\SE$. Note that the user's device
must be able to open a secure communications channel with $\SE$ and be
able to perform the cryptographic operations of Section
\ref{back:Brands}. The communication can be established in a standard
way with $TLS$. Denote $pk_{\USR}$, $pk_{\SE}$, $sk_{\USR}$,
$sk_{\SE}$, the public and private (secret) keys of $\USR$ and $\SE$
respectively. The TLS channel corresponding to these keys is denoted $TLS_{pk_{\USR},pk_{\SE}}$. Establishing these keys can be performed when $\USR$
sets up an account with $\SE$, independently of the enrollment steps of Section
\ref{sec:Enroll}. These steps should be anonymous for the user; namely it is not required that $\USR$ gives her real
identity to $\SE$ during this process, while $\SE$ should have a
well-known and trusted public key.

\section{Set-up phase} \label{sec:Setup} 

In order for service enablers to verify the authenticity of records issued by identity verifiers, we must begin with a set-up phase in which identity verifiers communicate to service enablers the Bitcoin address $a_{\IV}$ that they will use to issue new identity records in the enrollment phase, Section \ref{sec:Enroll}. $\IV$ should also communicate to $\SE$ the points $g_0\ldots g_n$ in secp256k1 that it will use for its selective disclosure protocols as in \ref{back:Brands}. 

\section{Enrollment phase} \label{sec:Enroll}

During the enrollment phase, $\IV$ will publish the root of a Merkle tree that commits to the (up-to-date) identities of all of its users $\USR_1,\ldots, \USR_N$. $\IV$ will update this information (at most) once per Bitcoin block, namely about once every 10 minutes, via a Bitcoin transaction that we will denote $\txenroll$.

\subsection{Calculation of DLREP} \label{sec:calc}

We detail the steps required for $\IV$ to calculate a DLREP encoding $\USR_i$'s identity. 

\begin{itemize}
	\item Each new user $\USR_i$ establishes her identity to $\IV$ by showing primary identity documents such as a passport of a driver's license. This is typical of the process of opening a bank account, for example. Denote the relevant user identity fields by $X_1,\ldots, X_n$.
	\item $\USR_i$ chooses a secret, random $X_{00}$, and communicates $h_{00}=g_0^{X_{00}}$ to $\IV$ along with a Brands proof that she knows an $X_{00}$ such that $h_{00}$ has the appropriate form (preventing an abusive user from submitting an $h_{00}$ of  an another form such as $ag_0^{b}$ for some chosen $a$ and $b$).
	\item $\IV$ chooses a random $X_{01}$, which is securely communicated to $\USR_i$. $\USR_i$ sets $X_0=X_{00}+X_{01}$. (We will see in Section \ref{sec:update} that this manner of collectively choosing $X_0$ between $\USR_i$ and $\IV$ will be useful when performing updates to issued identities.)
	\item $\IV$ computes $h_{\USR_i}=h_{00}\cdot g_0^{X_{01}}\cdot \prod\limits_{j=1}^{n} g_j^{X_j}=\prod\limits_{j=0}^{n} g_j^{X_j}$.
	
\end{itemize}

\subsection{Updating a DLREP and $X_0$} \label{sec:update}

%The enroll transaction $\txenroll$ of Section \ref{sec:Enroll} is protected from being altered by the immutability properties of the blockchain. On the other hand, we saw that new users are added via a subsequent $\txenroll$, which uses the output of the previous enroll transaction as an input. In this section we will detail how $\IV$ can also update some of the $X_i$'s for an existing user by using $\txenroll$. 

Sometimes a user's identity fields will change. She may change nationalities or she may turn 18 and no longer be marked as a minor. In this case, $\IV$ should have a convenient mechanism for updating the user's identity record which requires computing an updated $h'_{\USR_i}$. This presents the problem that, as we saw in Remark \ref{rem:dict}, $\IV$ should not publish two Brands commitments $h_{\USR_i}$ and $h'_{\USR_i}$ that share the same $X_0$. However, it may not always be practical for $\IV$ and $\USR_i$ to manually regenerate a new secret $X_0$ as in Section \ref{sec:calc}; there may be situations where $\IV$ is notified by governmental agencies such as tax or immigration services of a change in a user's identity, or a user may notify $\IV$ of a change in her identity in writing and may not be online to re-perform the exchange of $X_0$. Hence, $\IV$ should be able to unilaterally make changes to $\USR_i$'s records. 

%\subsubsection{Updating $X_0$}

We propose that for each update, the $X_0$ should be updated as well. Suppose a given $\USR_i$'s identity is being updated for the $k$-th time to commit to an identity of $X_1^{(k)},\ldots, X_n^{(k)}$. Then, $X_0$ will be replaced by $X^{(k)}_0=X_{00}+H^k(X_{01}),$ where $H^k$ is SHA-256 applied $k$ times. Even though $\IV$ does not know $X_{00}$, it can compute $g_0^{X^{(k)}_0}=h_{00}\cdot g_0^{H^k(X_{01})}$. Then, $\IV$ can compute the updated Brands commitment: $$h^{(k)}_{\USR_i}=h_{\USR_i}\cdot g_0^{X^{(k)}_0}\cdot \left(g_0^{X_0}\right)^{-1}\cdot \prod_{j=1}^{n}g_j^{X^{(k)}_j-X_j}.$$  %(For example, if $X_l^{(k)}=X_l$ remains the same after the update, $\IV$ does not need to know $X_l$ in order to compute this product.) %Namely, if only $X_l$ has changed to $X_l^{k}$ and $X_j=X_j^{(k)}$ for $j\neq k$,). 
 %which we will see below is sufficient to compute the updated $h'_{\USR_i}$. 

This schema has the result that \begin{itemize}\item If a field of
  $\USR_i$'s identity has changed, she can compute offline the new $X^{(k)}_0$,
  without having to re-perform an online communication  with $\IV$ to transmit $g^{X_0}$, as she knows
  both $X_{00}$ and $X_{01}$.  \item Only $\USR_i$ knows $X_{00}$,
  thus only $\USR_i$ has the information necessary to construct $X_0$
  and perform Brands proofs based on the commitment
  $h_{\USR_i}$. \end{itemize}

Furthermore, we want this process to preserve the same privacy guarantees as (static) Brands proofs. We argue that, in the random oracle model, for an attacker who does not possess $X_{01}$ or any of the $H^k(X_{01})$, the $X_{00}+H^k(X_{01})$ are randomly distributed (modulo the order of the elliptic curve). Hence, from the perspective of such an attacker, this process is equivalent to $\USR_i$ and $\IV$ re-performing the enrollment phase of Section \ref{sec:Enroll} to re-issue the identity with a new manually chosen, random $X_0$. See the appendix for more detail. 

%Specifically, if one attempts to perform the attack of Remark \ref{rem:dict} by taking versions of a user's identity $h^{(k_1)}_{\USR_i}=g_0^{H^{k_1}(X_{01})}\cdot \prod\limits_{j=1}^{n} g_j^{X_j}$ and $h^{(k_2)}_{\USR_i}=g_0^{H^{k_2}(X_{01})}\cdot \prod\limits_{j=1}^{n} g_j^{X'_j}$ that have been updated $k_1$ and $k_2$ many times respectively, one can compute $h^{(k_1)}_{\USR_i}/h^{(k_2)}_{\USR_i}=g_0^{H^{k_1}(X_{01})-H^{k_2}(X_{01})}\prod\limits_{j=1}^{n} g_j^{X_j-X'_j},$ but this is still blinded against dictionary attacks by the factor $g_0^{H^{k_1}(X_{01})-H^{k_2}(X_{01})}$. %but if the hash function $H$ is secure and $X_{01}$ is secret, this has a blinding factor against dictionary attacks of the same form as that of a (static) Brands commitment. 

%Now, if a field of $\USR_i$'s identity changes from $X_l$ to $X'_l$ $\IV$ can compute 
%$$h'_{\USR_i}=h_{\USR_i}\cdot g_0^{X'_0}\cdot \left(g_0^{X_0}\right)^{-1}\cdot g_l^{X'_l-X_l}.$$	Note that $\IV$ does not need to have know the values of the $X_j$ that do not change (namely for $j$ other than $l$). 

\begin{remark} 
	We recommend that at any given time, $\IV$ only store the $H^k(X_{01})$ necessary to perform the next update. As a result, even if $H^k(X_{01})$  is stolen from $\IV$'s servers, as long as this theft is detected before the user's identity is next updated, a thief will not have the means to remove the blinding factor and perform a dictionary attack against any of the previously published versions of the user's identity. Recall that these protections are based on the arguments of the appendix, which are themselves based on the work of Brands \cite{Brands}. %As these protections are based on the arguments of the appendix, which are based on the work of Brands \cite{Brands}. , they hold even the attacker also controls the Bitcoin infrastructure to which the identities are being published. 
\end{remark}

\begin{remark}
	Note that, in the computation of $h_{\USR_i}^{(k)}$, $\IV$ does not need to know the values of the $X_j$ that do not change, $X_j^{(k)}=X_j$. This can potentially allow identity verifiers to delete user information that they never expect to have to update later, even as the user can continue to their identity record to prove this aspect of their identity. As laws regulating the storage of personal data become more stringent, such as under the soon to come into effect General data protection regulation of the European Union \cite{euprivacy}, $\IV$ storing less customer data can mean less investment required in compliance departments and lower financial liabilities for data breaches. 
\end{remark}

\subsection{Publication of DLREPs}

$\IV$ will now post to the blockchain the root $r_{\IV,t}$ of a Merkle
tree that commits to the state at time $t$ of the identities for each
of its users. (Note that these are not the same Merkle trees that
exist natively in Bitcoin, see \ref{subsec:Bitcoinsyntax}, even though
the same data structure is used.) To compute $r_{\IV,t}$, $\IV$ must
both add new users to the tree and edit the information for existing
users if their DLREP has been updated as above. We describe how to
proceed. Standard trees are dynamic structures, where leaves
corresponding to an entry may move in the tree as its data evolves and grows in size. In
our case, we want to keep fixed the path to a user's entry, so she can
not authenticate using a previous, out-of-date, authentication path in
the tree.

As a consequence, we use a Merkle tree that is very large relative to the number of users to allow for future growth. Specifically, based on a simplified version of the data structure proposed in \cite{Coniks}, we use a virtual tree with $2^{256}$ leaves the paths to which are given using each bit of a SHA-256 hash to indicate which binary branch to take at each node. We denote $\USR$'s position in the tree by $\iota_{\USR_i}$, or the index of $\USR_i$ (see Figure \ref{fig:branch} for an example). In \cite{Coniks} this path is given by the hash of verifiable random function of a username; here, $\IV$ may attribute to each user an account number and then take the hash of this value, $\iota_{\USR_i}=H(\text{account number}_{\USR_i})$. $\IV$ will publish in an $\opreturn$ the root of this tree (at time $t$) $r_{\IV,t}$ and also a commitment that tracks all changes made to the tree $cm_{\IV,t}$ as below.
%As a consequence, we can only build new sub-trees for new users, and
%aggregate them with the previous tree, thus making a non-standard
%Merklized tree structure, which keep track of past users, as follows.  %Suppose $r_{\IV,t-1}$ is the
%previously published Merkle tree root.
	
	%Once $\IV$ has computed $h_{\USR_i}$
	
	\begin{itemize}	
	
	\item For each user $\USR_i$ of $\IV$, an index $\iota_{\USR}$ is calculated as above. $\IV$ creates a $256$ layer deep Merkle tree in which the leaf at each $\iota_{\USR_i}$ is $H(h_{\USR_i},\text{metadata}_{\USR_i})$, where $H$ is SHA-256. The metadata can include information such as an expiry data of the issued identity or other limitations on its use. All leaves that do not correspond to an $\iota_{\USR_i}$ contain an empty leaf. This tree has root $r_{\IV,t}$.
	
	\item If any updates have been made to pre-existing identities, then the leaf consisting of the user's (now updated) information should remain in the same location of the tree $\iota_{\USR_i}$. In particular, if the metadata has changed for an identity, such as the extension of its expiration date, etc, this can be appended to or replace the previous metadata.
	
	\item For the first commitment $\IV$ makes, it takes $cm_{\IV,1}=r_{\IV,1}$. Later, if $\IV$ has already computed $cm_{\IV,t-1}$, it computes $cm_{\IV,t}=H(cm_{\IV,t-1},r_{\IV,t})$.%versions of the previous roots: $r'_{\IV,1}=r'_{\IV, \text{new users},1}$, $r'_{\IV,l}=H(r_{\IV,l-1}, r'_{\IV,l-1},r'_{\IV,\text{new users},l})$ for $l=2,\ldots t-1$. ($r_{\IV,l-1}$ represents the original version as computed at time $l-1$; $r'_{\IV,l-1}$ represents the current version adjusting for updates.) 
	
	%\item Finally, $\IV$ computes $$r_{\IV,t}=H(r_{\IV,t-1},r'_{\IV,t-1},r_{\IV,\text{new users},t}),$$ which both appends the new users and takes into account the modifications to the existing users.
	
%	\item If $r_{\IV,\text{prev}}$ is the root published in the the previous $\txenroll$, $\IV$ computes $r_{\IV}=H(r_{\IV,\text{prev}},r_{IV,\text{new users}})$
	
	%\item If $rt_{\IV,\text{prev}}$ is the root published in the the previous $\txenroll$, $\IV$ computes $rt_{\IV}=H(rt_{\IV,\text{prev}},rt_{IV,\text{new users}})$. 
	
	\item $\IV$ publishes to the Bitcoin blockchain a transaction, $\txenroll$, of the following form:

%	{\footnotesize 
%	\begin{tikzpicture}[level distance=1cm,
%	level 1/.style={sibling distance=2.5cm},
%	level 2/.style={sibling distance=1.5cm}]
%	\tikzstyle{every node}=[circle,draw]
	
%	\node (Root)  {$r_{k}$}
%	child {
%		node {$r'_{k-1}$} 
%		child { node {} child{ node{$H(h_{\USR_1})$} } child{  }}%edge from parent node[left,draw=none] {help!} }
%		child { node {} child{  } child{ node{$H(h_{\USR_4})$} } }
%		%child { node {3} }
%	}
%	child {
%		node {$r_{new}$}
%		child { node {$H(h_{\USR_5})$} }
%		child { node{3} }
%	};
	
%	\end{tikzpicture} }
	%\begin{figure}{ \small 
	%		\centering 
			\begin{tikzpicture}
			\coordinate (up_left) at (-3,2);
			\coordinate (bottom_left) at (-3,0);
			\coordinate (up_middle) at (0,2);
			\coordinate (bottom_middle) at (0,0);
			\coordinate (bottom_right) at (5,0);
			\coordinate (up_right) at (5,2);
			\coordinate (label_left) at (-3,2);
			\coordinate (label_middle) at (0,2);
			\coordinate (label_right) at (5,2);

			\draw (bottom_left) -- (bottom_right) ;
			\draw (bottom_right) -- (up_right) ;
			\draw (up_left) -- (up_right) ;
			\draw (up_left) -- (bottom_left) ;
			\draw (up_middle) -- (bottom_middle) ;
			
			% decorations
			%\draw (up_left) node[above] {$\txenroll$};
		\draw (label_left) node[above right] {Input Addresses};
		%\draw (label_middle) node[left] {Amounts};
		\draw (label_middle) node[above right] {Output Addresses};
		%\draw (label_right) node[left] {Amounts};
		\draw (up_left) node[below right] {$\begin{array}{l}\addr_{\IV}\end{array}$};
		\draw (up_middle) node[below left] {$\begin{array}{r}\end{array}$};
		\draw (up_middle) node[below right] {$\begin{array}{l}\addr_{\IV}\\ \opreturn\left(r_{\IV,t}, cm_{\IV,t}\right)\end{array}$};
		%\draw (up_right) node[below left] {$\begin{array}{r}V\\ end{array}$};
	%	\draw (bottom_middle) node [above right] {Fees:};
	%	\draw (bottom_right) node [above left]{fee};
			
			\end{tikzpicture}
%		}
	%	\caption{Structure.}
		
	%	\label{fig:transaction0}

	%\end{figure}
	
	The input from the address $a_{\IV}$ should be the output of the previous $\txenroll$. One should calibrate to have one new transaction of this form in each block, namely approximately once every ten minutes. 
	
\end{itemize}

	As this transaction represents encoded semantic meaning in our system rather than financial data, the amounts of the inputs and outputs are secondary. Indeed, in the manner of \cite{Colu}, \cite{OpenAssets}, \cite{MIT}, the amounts should be at or near the minimum output amounts to be considered a standard transaction (\cite{MasteringBitcoin}, \cite{Dustcode}). 
	
	%The txid of the transaction in which $\USR_i$'s identity first appears, $\text{txid}_{\USR_i}$, should be communicated to $\USR_i$ along with the branch in the Merkle tree necessary to prove the presence of $H(h_{\USR_i},\text{metadata}_{\USR_i})$ as a leaf.
	%$\IV$ should save $h_{00}$ and $X_{01}$ for potential future updates to a user's identity (in fact, we will see that it is sufficient for $\IV$ to store $H(X_{01})$ where $H$ is the SHA-256 hash function, rather than $X_{01}$ itself, see Section \ref{sec:update}). However, $\IV$ does not necessarily need to maintain $X_1\ldots,X_n$ as $\USR_i$ will be able to use $h_{\USR_i}$ to prove statements about these fields even to $\IV$ if necessary. The minimizing of storing user data to what is strictly necessary is in keeping with the spirit of \cite{euprivacy}. 
\begin{remark}
	Note that the mechanism by which we create a Merkle tree of the $H(h_{\USR_i},\text{metadata}_{\USR_i})$ is similar to the structure of how \cite{Coniks} creates a public key registry, as implemented into a blockchain in \cite{Catena}. In \cite{Coniks}, $\iota_{\USR_i}$ is calculated with a keyed, verifiable function of a username in a way that is designed to prevent third parties from determining the position in the tree corresponding to a given username. This is done to prevent such third parties from tracking whether the information corresponding to that username has changed by tracking changes in the intermediate hashes along the branch to its location, which we see below must be circulated to other users that have indices near $\iota_{\USR_i}$. As $\iota_{\USR_i}$  in our case is not computed from public information, these precautions are less relevant but are nonetheless compatible with our system at the expense of additional overhead in the case that the policies of $\IV$ require them.

\end{remark}

\begin{remark}
	%By including the original value of $r_{IV,t-1}$ in the computation of $r_{IV,t-1}$ (in addition to the updated value), 
	Note that none of the history of the identity records emitted by $\IV$ can be changed without changing $cm_{\IV,t}$. As such, one could eventually migrate this system to another blockchain by simply transferring the last value $cm_{\IV,t}$, which ensures the integrity of all of the preceding information.  
	
\end{remark}

\begin{remark}
	Rather than forming a hash tree of the identity records of $\IV$'s users, it would be more straightforward for each user to have their $h_{\USR_i}$ included in an $\opreturn$ in a separate Bitcoin transaction. This is the model for $\cite{MIT}$, where each certificate is issued via its own transaction. Unfortunately, this presents scaling problems. First, as the number of transaction which $\IV$ has to issue increases, this increases the costs paid in transaction fees (compare to our estimates of these costs in Section \ref{sec:cost}). Second, Bitcoin blocks are currently limited to 1MB in total size and calibrated so that there is one block per 10 minutes \cite{MasteringBitcoin}. As a result, there is a limit on the total number of Bitcoin transactions that can issued of about 7 transactions per second \cite{Blockstack}. While there are efforts to increase this limit \cite{Segwit}, our system if widely adopted risks having a number of users that exceeds what could be supported by the network if each must have her own issuing transaction.   %is currently a limited amount of total bandwidth in the Bitcoin system (see \cite[Section IV-B]{Catena}).  	
\end{remark}

\subsection{Information to be distributed by $\IV$} \label{subsec:distributed}

In order for a user $\USR_i$ to be able to authenticate against her identity $h_{\USR_i}$, she will need to be able to demonstrate that $h_{\USR_i}$ is present in the last published Merkle tree. For this, the user will need the txid of one of $\IV$'s issuing transaction which can then be traced to the most up-to-date information published by $\IV$, $\text{txid}_{\IV}$, the location of her entry in the Merkle tree, $\iota_{\USR_i}$, and the intermediate hashes of the tree along the branch where this entry is located. If $\USR_i$'s identity is updated, in the most recent $r_{\IV,t}$ this branch will lead to the updated commitment $cm_{\IV,t}$, so $\USR_i$ will no longer be able to authenticate her old identity against it. 

$\USR_i$ will need to know if her identity is updated. We imagine that if some field of $\USR_i$'s identity has been invalidated by $\IV$, that $\IV$ must notify her. Thus, $\USR_i$ should have up-to-date values of her $X_1\ldots X_n$. She can track the number of notifications of updates $k$ that she receives in order to be able to calculate the current value of $X^{(k)}_0=X_{00}+H^k(X_{01})$.

We also envisage $\IV$ distributing non-sensitive information to the various service enablers that is nevertheless important to the functioning of the system. It makes sense for this information to be stored by $\SE$ whose primary role, in contrast to $\IV$, is in the facilitation of the identity system. Thus, we can avoid $\IV$ having to be
``lively'' and ready to distribute information at any given time to various users. As these $\SE$ \textit{are} required by their business model to be lively so as to be able to enable authentications, this additional role does not substantially burden them.

Specifically, if any user's entry in the Merkle tree is updated, that will affect the intermediate hashes that then need to be made available to the other users a service which can be provided by $\SE$. Hence, we imagine $\SE$ storing the tree information $(H(h_{\USR_i},\text{metadata}_{\USR_i}),\iota_{\USR_i})$ so that these intermediate hashes can be recalculated as necessary. See Section \ref{sec:bandwidth} for estimation on how much data must be stored. On the other hand, if $\IV$ distributes all this information to all of the $\SE$ in the market, this results in a substantial redundancy of a significant amount of data. Hence, the best course of action may be to have $\IV$ distribute the entire tree to a small number of service enablers, then when a user wants to use her identity, she can download the required information from one of these service enablers whether that is the $\SE$ that she will later correspond with in the proof of identity phase (see Section \ref{sec:Auth}) or not.  

\subsection{Comments on updating and revocation}

As a special case of updating, $\IV$ can revoke an identity by replacing $h_{\USR_i}$ with an empty string. 

$\SE$ can be assured of having the most recent information published by $\IV$, because if $\SE$ checks an out-of-date entry, it will see that the output to $\addr_{IV}$ has been spent. $\SE$  can then follow a chain of transactions to the last published Merkle root. 

Our system takes advantage of the fact that, in Bitcoin, a decentralized, Byzantine fault resistant, highly available system of nodes is tracking whether updates are made to their list of UTXOs. As any update made to our system implies an update tracked by these nodes, this provides an effective mechanism to track updates to our system.  See \cite{MIT} and \cite{Catena} for architectures that employ a similar idea in different use cases.

\section{Proof of identity phase } \label{sec:Auth}

In this section we show how $\USR$ can use an identity that she has
been issued by $\IV$ via the process of Section \ref{sec:Enroll} to
authenticate herself and gain access to a service provided by $\SP$. We suppose that $\USR$ has downloaded the intermediate hashes on the path to her entry of the Merkle tree as in Section \ref{subsec:distributed} from some service enabler (that is not necessary the same as $\SE$ below).
As we discussed in Section \ref{sec:actors}, $\USR$ and $\SE$ should have previously properly set-up the necessary keys to open the secure communications channel $TLS_{pk_{\USR},pk_{\SE}}$. (This might occur upon $\USR$ downloading $\SE$'s mobile app and opening an account). Then, the following steps are performed, where communication between $\USR$ and $\SE$ is done through this channel:

\begin{algorithm}
	\begin{algorithmic}[1]
		\renewcommand{\algorithmicrequire}{\textbf{User data:}}
		\renewcommand{\algorithmicensure}{\textbf{Output:}}
		\REQUIRE  $pk_{\USR}$, $sk_{\USR}$, $pk_{\SE}$, $\text{txid}_{\IV}$, $\iota_{\USR}$, branch of Merkle tree, $X_0,\ldots X_n$.  %via $TLS_{pk_{\USR},pk_{\SE}}$
		%\ENSURE  out

		\STATE Request $\USR\rightarrow\SE$:
                $(\text{Name of service, }\SP)$

                \STATE
                Determine what to prove: $\SE \leftrightarrow
                \SP$: 
                \begin{itemize} 
                \item $\SE\rightarrow\SP$:
                  $(\text{Name of service}, \text{ Session id})$
                
                \item $\SP\rightarrow\SE$: $(\text{Info to prove}$, $\{\IV_1,\ldots, \IV_m \},$ \\ $\text{Session id})$

                \end{itemize}

		\STATE Prove: $\USR \leftrightarrow \SE$:

		\begin{itemize} 

                \item $\SE\rightarrow\USR$: $(\text{Info to prove, }
                  \{\IV_1,\ldots, \IV_m \},$ \\ $\text{Session id})$
                \item $\USR\rightarrow\SE$:  $(\text{txid}_{\IV_j}\text{, }h_{\USR},  \text{ branch of
                  Merkle tree},$ \\ $\text{Session id})$
                \item $\SE$ uses the provided branch and the publicly available, most recent version of $r_{\IV,t}$ to check that $h_{\USR}$ is present in the tree and up-to-date
                  
                \item $\USR\leftrightarrow\SE$:
                  (interactive proof $\pi$, Session id)

                \item $\SE$ checks the proof%, and that $h_{\USR}$ is not revoked/up to
              %    date 
              \end{itemize}
		
		\STATE Confirm: $\SE\rightarrow\SP$: (Session id) 
		\STATE Grant of service: $\SP\rightarrow\USR$ (if this service is digital, it might pass through the established connections between $\SP$ and $\SE$ en route to $\USR$)
	
	%	\RETURN 
	\end{algorithmic} 
\end{algorithm}

\begin{figure} 

	\begin{tikzpicture}[decoration={
	markings,% switch on markings
	mark=% actually add a mark
	at position 2cm
	with
	{
		\draw (-2pt,2pt) -- (2pt,0pt);
		\draw (-2pt,-2pt) -- (2pt,0pt);
	}
}
]*2/3

\coordinate (up_middlel) at (-.35,2);
\coordinate (up_middler) at (.3,2);
\coordinate (up_middle) at (0,2);

\coordinate (bottom_left) at (-3,0);
\coordinate (bottom_leftu) at (-2.8,.3);
\coordinate (bottom_leftd) at (-2.8,-.3);

\coordinate (bottom_right) at (3,0);
\coordinate (bottom_rightu) at (2.8,.3);
\coordinate (bottom_rightd) at (2.8,-.3);

%\coordinate (bottom_middle) at (0,0);
%\coordinate (bottom_nearmiddle) at (.25,-.1);
%\coordinate (below_left) at (-3,-1);
%\coordinate (below_right) at (3,-1);

%\coordinate (far_below_left) at (-3,-2);
%\coordinate (far_below_right) at (3,-2);

% entities
\draw (up_middle) node[above] {$\SE$};
\draw (bottom_left) node[left] {$\USR$};
\draw (bottom_right) node[right] {$\SP$};
\draw  [postaction={decorate}] (up_middler)  to [bend left=30] node[sloped,midway,above] {2. Determine what to prove}(bottom_rightu);% 
\draw  [postaction={decorate}] (bottom_rightu)  to [bend right=30] node[sloped,midway,above] {} (up_middler);% 
\draw  [postaction={decorate}]  (up_middlel)  to [bend right=30] node[sloped,midway,above] {3. Brands proofs}(bottom_leftu);% 
\draw  [postaction={decorate}] (bottom_leftu)  to [bend left=30] node[sloped,midway,above] {} (up_middlel) ;% 
\draw  [->] (up_middlel)  to [bend right=30] node[sloped,midway,above] {}(bottom_leftu);% 
\draw  [postaction={decorate}]  (up_middler)  to [bend right=15] node[sloped,midway,above] {4. Confirm}(bottom_rightu);
%\draw [postaction={decorate}] (up_middler) to[bend left] (bottom_rightu) node[midway,above,sloped] {4. Confirmation} ;
\draw [postaction={decorate}] (bottom_rightu) -- (bottom_leftu) node[midway,below,sloped] {5. Grant service} ;
\draw [postaction={decorate}] (bottom_leftu) to [bend right=15]  node[sloped,midway,above] {1. Request} (up_middlel);

\end{tikzpicture}
\caption{Schema of interactions between $\USR$, $\SE$, and $\SP$ during the proof of identity phase.}
\end{figure}
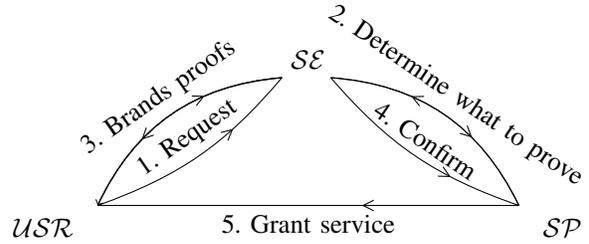 

\begin{figure}
	\begin{center} 
	\tikzset{
		treenode/.style = {shape=rectangle, rounded corners,
			draw, align=center,
			top color=white, bottom color=blue!20},
		root/.style     = {treenode, font=\Large, bottom color=red!30},
		env/.style      = {treenode, font=\ttfamily\normalsize},
		dummy/.style    = {}
	}
	\begin{tikzpicture}
	[
	edge from parent/.style = {draw, -latex},
	every node/.style       = {font=\footnotesize},
	sloped
	]
	\node [dummy] {Computed $\stackrel{?}=r_{\IV,t}$}
	child { node [dummy] {Provided}
		edge from parent node [above] {0} }
	child { node [dummy] {Computed}
		child { node [dummy] {Computed}
			child { node [dummy] {$h_{\USR}$}
				edge from parent node [above] {0}
			}
			child { node [dummy] {Provided}
				edge from parent node [above] {1} }
			edge from parent node [above] {0} }
		child { node [dummy] {Provided}
			edge from parent node [above, align=center]
			{1}
		}
		edge from parent node [above] {1} };
	\end{tikzpicture}

	\end{center} 
 \caption{The verifications performed by $\SE$ to prove that $h_{\USR}$ is a branch in the tree using the intermediate hashes of the branch provided by $\USR$. Namely, at each step moving up from the bottom, $\SE$ computes the hash of what has already been computed with the next provided intermediate hash and compares the ultimate result with $r_{\IV,t}$. Note that $\iota_{\USR}$ in this example, which gives the position of $h_{\USR}$ in the tree, is $100$.}
 	\label{fig:branch}
 
\end{figure}
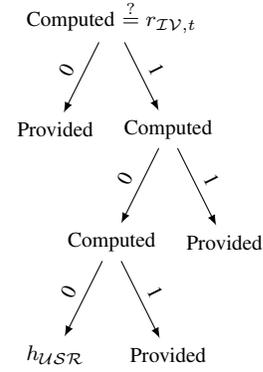

In more detail, over the TLS channel between $\USR$ and $\SE$, $\USR$ communicates to $\SE$ what service she is requesting from which service provider.
%in order to authenticate, $\USR$ communicates to $\SE$ (via a website or a mobile app) $pk_{\USR}$, and which service she is requesting from which service provider. To exchange this information, $\USR$ and $\SE$ use their corresponding keys to open a TLS channel. 
Then, $\SE$ creates a session id and then communicates with $\SP$ to establish %the sends it to $\SP$ along with $pk_{\SE}$ to $\SP$. Also, $\USR$ signs a message to establish to $\SP$ that she does, in fact, control the corresponding private key. 
%Then, upon $\SP$ querying $\SE$ for the validity of $\USR$'s identity by indicating 
what information about $\USR$'s identity needs to be established and provides a list $\{\IV_1,\ldots, \IV_m \}$ of identity verfiers from which the service provider accepts documents. % $\IV$'s that service provider accepts documents from. %$\USR$ and $\SE$ use their corresponding keys to open a TLS channel. 
Again over the TLS channel between $\USR$ and $\SE$, $\USR$ communicates the txid, $h_{\USR}$, and the necessary branch of the Merkle tree corresponding to her relevant identity as established in Section \ref{sec:Enroll}. %Thus, $\SE$ has access to the Brands commitment $h_{\USR}$ 
Using the txid, $\SE$ can follow the chain of updates, as in Section \ref{sec:update}, to obtain the most up-to-date version of $r_{\IV,t}$. Then, $\SE$ can use $h_{\USR}$ and the intermediate hashes on the branch to compute a value that can be compared to $r_{\IV,t}$ to confirm the presence of $h_{\USR}$ in the most up-to-date version of the tree, see Figure \ref{fig:branch}.
% this identity and can follow the chain of updates to this identity to verify if it is revoked or otherwise updated as in Section \ref{sec:update}. 
$\SE$ can also verify that this identity was issued by an $\IV$ on the list provided by $\SP$. $\SE$ makes interactive requests of knowledge which $\USR$ establishes via the interactive proof $\pi$. Note that the ability of $\USR$ to provide these proofs depends on her knowledge of $X_0,\dots X_n$. Particularly, $X_0$, which only $\USR$ should know as she randomly generates the secret $X_{00}$ component, acts as a sort of key to the use of this identity record. 

Once $\SE$ is satisfied with the proofs provided by $\USR$, $\SE$ sends a confirmation to $\SP$, and $\SP$ grants the service to $\USR$.

\begin{remark}
	While the Brands proofs protect the user from having to reveal unnecessary identity information to $\SP$, $\SE$ nevertheless observes details about $\USR$'s transactions, specifically with which service providers $\USR$ is interacting. If $\SE$ notices suspicious behavior on the part of the user, $\SE$ might refuse to process a transaction or contact $\IV$ with a warning in a manner akin to how fraudulent credit card activity is detected. 
\end{remark}

\begin{remark}
	Note that, if $\SP$ is not willing to completely trust any single service enabler, a user might be required to perform this process with several different $\SE$. A smart phone app with which the user has already exchanged keys to establish TLS connections with each $\SE$ could effectively coordinate this process. 
	
\end{remark}

\section{Key loss}

We have seen that users must store the following information: $pk_{\USR}$, $sk_{\USR}$, $pk_{\SE}$, $\text{txid}_{\IV}$, and $X_0,\ldots X_n$ in order to use to perform authentications (and she must also have the intermediate hashes on the branch of the Merkle tree on which their identity record is located, possibly re-downloading them from a service enabler). We consider the contingencies if a user loses access to some or all of this information due, for example, to a lost phone. 

Note that $sk_{\USR}$ serves only to establish a secure communications
with $\SE$, allowing them to exchange (potentially) sensitive
information. Particularly, as this key corresponds to an account that
can be obtained anonymously, $sk_{\USR}$ is not the basis for proving
user identity, and serves mainly to protect against man-in-the-middle
attacks, where the attacker intercepts and replays messages from $\USR$
in order to be granted access to $\SP$. This has the consequence that
if $sk_{\USR}$ is lost, $\USR$'s identity is not compromised, nor is her ability to perform Brands proofs. Hence, in this case, $\USR$ should simply generate a new pair
$sk'_{\USR}$, $pk'_{\USR}$, and communicate $pk'_{\USR}$ to $\SE$.

On the other hand, if $X_0$ is lost, and there is not a backup available to the user, she should contact $\IV$, again with her physical identity documents to justify her identity. Then $\IV$ can issue an update to an existing identity record to establish a new $X_0'$ as in Section \ref{sec:update}.

\section{Security model} \label{sec:security}

As we discussed in Section \ref{sec:actors}, $\USR$ must trust both $\IV$ and $\SE$ to handle her personal information appropriately; however, the information that $\USR$ must convey to $\SE$ is generally much more limited than that conveyed to $\IV$. Technically, 
$\USR$ and $\SE$ should have the capacity to perform interactive Brands proofs. As we saw in Proposition \ref{prop:Brands}, the exchange of Brands proofs reveals nothing about an identity except what statements are being proved. Nonetheless, these statements may be themselves sensitive.  By encrypting them, our system protects this information in the same matter as a tradition online authentication system even as the information that the user must provide to $\SE$ is minimized. %Note that $\USR$ must trust $\SE$ to not act as a man-in-the-middle and use the proofs conveyed to it to impersonate the user. 

%Users need to trust identity verifiers to use their personal information in a way that respects user privacy (just as an individual needs to trust their electric company or bank to not misuse their information). Meanwhile, service enablers perform a role of being trustworthy relays to service providers. As such, $\SP$ is trusting $\SE$ to not lie about given user authentications. While $\USR$ conveys much less sensitive information to $\SE$ than to $\IV$, $\SE$ should nonetheless handle this information in a responsible manner. Particularly, one must trust $\SE$ to not act as a man-in-the-middle and use the proofs conveyed to it to impersonate the user. 

We assume that $\IV$ properly issues updates to an identity $h_{\USR}$ by overwriting the previous version of that identity at index $\iota_{\USR}$. If $\IV$ fails to do this $\USR$ may have the opportunity to continue to authenticate herself against an identity that is out-of-date but appears valid to $\SE$. Note that if $\IV$ realizes that this type of error has been made (either due to coding error, infiltration by hackers, etc), then it can correct these errors in the next update, revoking redundant entries. 

Abstractly, we make use of the \emph{public ledger} functionality of Bitcoin, namely that it acts
as a ``bulletin board'' on which anyone can post messages and read the messages that have previously been
posted. Specifically, we require that Bitcoin have the properties of  \emph{liveness}, i.e. every honest
participant will have its posted messages seen by every honest
participant after some delay, and \emph{persistence}, namely that
every posted message will indefinitely be seen at the same position by
all participants, see \cite{GarayKL15} and \cite{PassSS17}. We also depend on the integrity of the Bitcoin transaction verification procedure to check the validity of transactions which includes checking that each non-generation transaction has inputs
corresponding to previous transaction outputs, etc. The Bitcoin core protocol has been proven to have these properties when quantitative bounds are assumed on the relative power of the adversary to the honest players, in terms of computing power
in~\cite{GarayKL15}, or in terms of computing power and influence over the peer-to-peer network in
~\cite{PassSS17}. However, these results are theoretical; in addition to the risk that a hostile adversary may become powerful enough to invalidate the quantitative bounds, there is also the possibility that coding bugs or other accidental situations may cause problems such as small forks, peer-to-peer
failures, etc \cite{longfork}. 

More concretely, the Bitcoin aspect of our protocol serves to allow $\IV$ to update an identity in an unequivocable manner, and provides neutral infrastructure that is computationally protected against unauthorized modifications. The consequence for our system of a fork would be to allow a user to continue to authenticate against a non-updated copy of an identity that should have been updated. This can be particularly consequential if $\IV$ attempts to revoke a user identity. As a result, service providers and service enablers might consider suspending particularly sensitive authentications if $\SE$ detects indications of a fork.

 Note that the service enabler is the only actor in our schema that needs to be capable of verifying the status of Bitcoin transactions. In order to obtain the most up-to-date information on the network available, $\SE$ should operate a full node. (However, a user who wants to emulate the checks performed by $\SE$ to see whether her identity is still valid could obtain information about the Bitcoin network from running a Simplified Payment Verification client \cite{MasteringBitcoin} or even one or more block explorers, accepting the risk that attacks against these methods might hide from her an update to her identity that would be seen by a full node.)

It is common practice for merchants accepting bitcoin to wait several confirmations (several mined blocks) before accepting a transaction as valid. A common rule-of-thumb is to wait six confirmations (approximately one hour) after a transaction to be confident that this transaction cannot be reversed \cite[Chapter 2]{MasteringBitcoin}. In our scheme, the only party who can double spend is $\IV$, who we do not generally expect to issue malicious, contradictory updates. If $\IV$ issues an update transaction that is included in an orphaned block, but ultimately does not make it into the main chain, $\IV$ can merely include those user updates in a reissued transaction later. 

$\SE$ and $\SP$ may still want to wait for a transaction to be
included in a few blocks so as to avoid authenticating users against
ephemeral records for auditing reasons. However, since only $\IV$ can
perform double spending in our scheme, our
system does not require the same level of caution against double
spending attacks that is employed for substantial financial
transactions in Bitcoin.

Finally, as Bitcoin is integrated into our structure, a network attack that makes it impossible for $\IV$ to issue new transactions or for $\SE$ to obtain information on the latest transactions would function as a denial of service attack on our system. This reflects the importance of using a well-established blockchain such as Bitcoin with a large network rather than a smaller, newer blockchain to be as robust as possible against such attacks.  %$\SE$, as the only actor in our schema that must verify the status of Bitcoin transactions, should obtain the most up-to-date information on the network available, by operating a full node. 

\section{Bitcoin transactions costs} \label{sec:cost}

We estimate the cost in Bitcoin transaction fees for an $\IV$ to use this system. As the $\txenroll$ transaction has one $\opreturn$ output containing two SHA-256 hashes, P2PKH output to $a_{\IV}$, ---and one input (corresponding to a  P2PKH output of a previous transaction), the total size of its raw transaction is approximately $265$ bytes \cite{MasteringBitcoin}. The amount of bitcoin that must be paid in fees for a transaction of a given sizes fluctuates based on market forces as miners must choose which transactions to include in the block they are mining with limited space; current (May 2017) estimates \cite{fees} suggest that a fee of .0000036 bitcoin per byte is sufficient to have a high likelihood that a transaction will be included in the next block. This results in a fee for $\txenroll$ of approximately .000954 bitcoin or 2.17 USD (1 BTC=2280 USD, May 2017). If $\IV$ has enough clients to justify emitting a $\txenroll$ every 10 minutes, this would require transaction fees of approximately 312 USD per day.

\section{Storage and bandwidth requirements} \label{sec:bandwidth}

%Second, we consider the bandwith requirements when users need to get
%information about updates of their authentication path in the Merke
%tree. Consider that $\IV$ manages $N=2^{23}$ users. The dynamics of
%updates are low since user identity is rather static, so for example,
%we consider that a fraction $f_{\text{daily}}=1\%$ of users have their
%identity fields changing daily. This means a fraction
%$f= f_{\text{daily}}/(6\cdot 24)$ updates every Bitcoin block. 
%At time
%$t$, assuming all previous batches of new users with Merkle root
%$r_{\IV,\text{new users}, l}$, $l=1\dots t$, have the same size
%$\nabla$, we have $\nabla=N/t$. If these updates are distributed
%uniformly, we have an expectancy of $fN/\nabla=ft$ updates in each
%batch. Following similar arguments to the ones done by
%% CONIKS~\cite{Coniks}, this implies an average of $\log_2(ft)+t-l+1$ hashes
%to be communicated $\USR$ supposing her data has been embedded at time
%$l$ in the Merkle tree. For our example, this gives $\log(f)+\log(t)+t-l+1$ hashes, with $t\leq N/(24\cdot 6)=$

%Note: this section assumes the $h_{\USR}$ are stored in the leaves of a  symmetric Merkle tree similar to that used by \cite{Coniks}. Namely, an (excessively large) tree with $2^{256}$ leaves the paths to which given by a SHA-256 hash of user information. (Not the current model proposed in this paper.) 

Consider that $\IV$ manages $N=2^{26}\approx 67$ million users, the size of a large bank. The dynamics of
updates are low since user identity is rather static, so for example,
we consider that a fraction $f_{\text{daily}}=1\%$ of users have their
identity fields changing daily. This means a fraction
$f= f_{\text{daily}}/(6\cdot 24)$ updates every Bitcoin block.

$\IV$ will issue updates which will be reflected in the root hash of the Merkle tree created in the enrollment phase. However, as discussed in Section \ref{subsec:distributed}, during the interaction between $\SE$ and $\USR$ during the proof of identity phase, these parties need to have access to the intermediate hashes of this Merkle tree so that they can verify that $h_{\USR}$ is present in the most recent version of the tree.  %We present and compare two options for the management and distribution of these intermediate hashes.

\subsection{$\SE$ stores information on behalf of $\IV$}
\label{subsec:storage}
In order to be able to calculate the appropriate intermediate hashes for any potential user of $\IV$'s system, $\SE$ needs to store all of the (occupied) leaves of the tree. This means storing, for each leaf, a Brand commitment and the index of this leaf in the tree $(h_{\USR}, \iota_{\USR})$, which consists of a total of $64$ bytes per leaf. Hence, $\SE$ must store $N\cdot 64$ bytes, or under our assumptions  $2^{26}\cdot 64=4.3$ gigabytes for each $\IV$ for which this service is provided. In an average block then $\SE$ will need to update $f\cdot N\cdot 64$ bytes, approximately $298$ kilobytes each $10$ minutes for each $\IV$. 

Additionally, in order to maintain a full node, $\SE$ must download all new blocks, namely approximately 1 MB every $10$ minutes, independent of the number of identity verifiers.

\subsection{$\USR$ stores information necessary for her own authentications}

For $\USR$ to store the intermediate hashes with which she can verify her own identity against the root hash of the tree, she does not need to store the entire tree. Instead, she can merely store the single path down the tree against which her identity will be hashed, to be presented to $\SE$ during the proof of identity phase. A priori, this would require storing $256$ hashes each of $32$ bytes, each of which will have to be regularly updated as the other users update their information. However, as most of the leaves of the tree are empty, in fact, $\USR$ will on have average lower storage and bandwidth requirements.  

Based on the same argument as in \cite[Section 5.3]{Coniks}, an average user will only have $\log_2(N)$ many non-empty intermediate hashes on her authentication path. If for each intermediate hash, $\USR$ also stores its location in the tree, this corresponds to $64 \log_2(N)$ bytes, or under our assumptions to storage of $1664$ bytes. Similarly, as there are $f\cdot N$ modification made to the tree per block, on average only $\log_2(f\cdot N)$ elements of $\USR$'s authentication path in the tree will change. %intutive argument: fN of changes affect root, fN/2 of changes affect each intermediate hash of next stage up, fN/4 of changes affect each intermediate hash of next step up, etc. When get log(fN) steps up this is less than one, most intermediate hashes not affected/don't need to be changed.
For each change, $\USR$ requires the new value of this intermediate hash and its position in the tree. Namely, $\USR$ must download $64\log_2(f\cdot N)$ bytes of information for a total of approximately $780$ bytes each block under our assumptions.

Thus, when a user wants to use her identity, she can download the required information from one of these $\SE$ whether that is the service enabler that she will correspond with in the proof of identity phase or not. Depending on how much time has passed since the user was last online, she may need to download between $780$ and $1664$ bytes as discussed in Section \ref{subsec:storage}. 

\section{Comparison of revocation mechanism to revocation in PKIs}

We briefly consider the advantages and disadvantages of the update and revocation mechanism presented in Section \ref{sec:Enroll} to the standard revocation systems that are used in public key infrastructures: certificate revocation lists (CRLs) and the Online Certificate Status Protocol (OCSP). 

Even though use of CRLs is limited to revoking certificates, rather than allowing for more complex updates, their size, which is linear in the number of revocations, can become quite large. The median certificate has a revocation list on the order of 50 KB, and CRLs of several megabytes are not uncommon \cite{revocationlists}. 
We saw in Section \ref{sec:bandwidth} that the amount of information required to verify the status of a user's identity against an up-to-date $r_{\IV,t}$ is logarithmic in the number of $\IV's$ users, on the order of 1 KB. On the other hand, as also discussed in Section \ref{sec:bandwidth}, the bandwidth required to run a Bitcoin full node to be able to confirm that $r_{\IV,t}$ is, in fact, up-to-date is roughly 1MB every ten minutes, independent of the number of identity verifiers. There are nonetheless situations where $\SE$ is interacting with many identity verifiers, particularly when $\SE$ is performing the role of a single sign on service, where our update model would be advantageous from a bandwidth perspective. Moreover, a user that runs a Simplified Payment Verification client to validate that her identity is up-to-date, need only download 80 byte block headers every ten minutes and approximately 1 KB of information for each verification of $r_{\IV,t}$ \cite{MasteringBitcoin}.

OCSP was developed to alleviate the bandwidth concerns of CRLs at the expense of requiring Certificate Authorities to be ``lively'' in responding to requests on revocation status in real time. As discussed in Section \ref{sec:actors}, we seek to avoid requiring such liveliness from identity verifiers.

%Moreover, 

%The tradeoff here is that  information  \ref{sec:security} that 

%traditional architectures where users are empowered to manage the day-to-day use of their identities without direct interaction with identity providers, such as revocations lists in PKI's.

Moreover, as shown in \cite{revocationlists}, a number of widely used, modern browsers accept revoked certificates in certain circumstances. In contrast, note that our update mechanism is integrated into the issuing system so that checking that a record is up-to-date is done in the same process as checking that that record exists at all. Hence, we minimize the possibility for authentication to be separated from appropriate revocation.

Finally, denial of service attacks can be performed on public key infrastructures by preventing verifiers from downloading CRLs or communicating with a OCSP server. In our system, as we saw in Section \ref{sec:security}, such attacks are defended by the infrastructure of the Bitcoin network. In exchange for the protections of this network, potentially non-negligible Bitcoin fees must be paid, as seen in Section \ref{sec:cost}.%, and making use of a $\SE$ who maintains up-to-date information about the Bitcoin blockchain, see Section \ref{sec:security}.

As such, while we are not proposing to completely replace PKI models of revocation with blockchain based models (indeed, the TLS connections we propose using during the proof of identity phase make use of traditional public key infrastructures), we argue that the revocation and update mechanisms we have presented are a useful tool that will be beneficial in certain use cases.

\section{Conclusion}

We have presented an architecture in which users can authenticate themselves against records established by ``identity verifiers,'' such as a bank or a utility company, even as these verifiers do not need to store the personal data of the user directly, following the spirit of \cite{euprivacy}. %, such as banks and electric companies, who are trusted by service providers as authenticating individuals ....?
As the business model of the verifers does not necessarily focus on identification, we outsource much of the operation of this system to a third party ``service enabler'' who nonetheless learns no more than what is necessary about the identity of the user. By making use of the Bitcoin blockchain, we allow for a verifier to update or revoke a user's identity in a streamlined fashion.

%We will discuss the impact of these possible attacks
%and failures in Section~\ref{real-world-security} below.

%$\SE$ should have the capacity to perform interactive Brands proofs and to verify the state of a given Bitcoin transaction such as by operating a full node. 

%\section{Potential attacks}

%\section{Example use case}

\bibliographystyle{IEEEtran}
\bibliography{acgpst}

\appendix

\section{Securiy of the update method in section~\ref{sec:update}} \label{appen}

We use the framework of Brands' thesis~\cite{Brands} to discuss and
analyze the security of our update method in~\ref{sec:update}. In
particular, \cite[Definition 2.3.1]{Brands} introduces the notion of
\emph{instance generators} for one-way and collision intractable
functions. In the DLREP case, an instance generator is a way to
generate the group and the $g_i$'s, and also to generate
$X_0,\dots,X_n$. In Section~\ref{back:Brands}, we have essentially
recalled Brands standard DLREP instance generator, and as soon as
$X_0$ is random, the DLREP function is one-way and collision
intractable \cite[Proposition 2.3.3]{Brands}. As noted by
Brands~\cite[page 61]{Brands}, if an instance generator is
indistinguishable from the basic DLREP instance generator, the
security theorems of his thesis still hold.

We prove here that, under the random oracle model (i.e.\ the hash
function $H$ is a random function), the instance generator provided by the 
method in ~\ref{sec:update} is indistinguishable from the basic method
in~\ref{back:Brands}. In~\ref{back:Brands}, $X_0$ is generated at
random, while in~\ref{sec:update}, $X_0^{(k)}=X_{00}+H^k(X_{01})$, where
$X_{00}$ and $X_{01}$ are random. Since $H$ is modeled as a random
function, and since the addition is done in a cyclic group, $X_0^{(k)}$ is
as random as $X_0$ in the basic method, thus, the instance generators
are statistically indistinguishable. This proves that each updated
$h^{(k)}_{\USR_i}$ is a one-way and collision intractable function of
$X_{00}$, $X_{01}$, and $X^{(k)}_1,\dots,X^{(k)}_n$.

Now we discuss the security with respect to a malicious
$\IV$. \cite[Proposition 2.4.8]{Brands} and the following discussion
also shows that the basic protocol for proving knowledge of a DLREP
in~\ref{back:Brands} is complete and sound for the basic instance
generator. Being indistinguishable, the same holds true for our instance
generator in~\ref{sec:update}. Thus, if a dishonest $\IV$, knowing
$X_{01}$, wants to use $h^{(k)}_{\USR_i}$ to authenticate as the user, it
has to know $X_0^{(k)}=X_{00}+H^k(X_{01})$ (completeness), which is equivalent to  $\IV$ knowing $X_{00}$ (subtraction). Thus, it is not feasible for $\IV$ to authenticate without  %if $X_{00}$ is
%secret to $\IV$,  
knowing the secret $X_{00}$, which is protected by the hardness of the discrete logarithm problem. %The same holds when considering consecutive $h'_{\USR_i}$.
This continues to hold for each successive updated $h^{(k)}_{\USR_i}$. 
%\nocite{*}
%\IEEEtriggeratref{8}
%\bibliographystyle{IEEEtran/bibtex/IEEEtran}
%\bibliographystyle{IEEEtran}
%\bibliography{acg-pst}
%\section{Example}
%\begin{center}

%\end{center}
\end{document}